\documentclass[a4paper,fleqn]{cas-dc}

\usepackage{amssymb}
\usepackage{graphicx}
\usepackage{float}
\usepackage{amsmath}
\usepackage{enumitem}
\usepackage{bm}
\usepackage{multirow}
\usepackage{colortbl}
\usepackage{xcolor}
\usepackage{hyperref}

\usepackage[subtle]{savetrees}
\newcommand{\ie}{\emph{i.e., }}
\newcommand{\eg}{\emph{e.g., }}

\newcommand{\wrt}{\emph{w.r.t. }}

\usepackage[numbers]{natbib}
\def\tsc#1{\csdef{#1}{\textsc{\lowercase{#1}}\xspace}}
\tsc{WGM}
\tsc{QE}

\begin{document}
\let\WriteBookmarks\relax
\def\floatpagepagefraction{1}
\def\textpagefraction{.001}

\shorttitle{Proactive Recommendation in Social Networks: Steering User Interest with Causal Inference}    

\shortauthors{Hang Pan et al.}  

\title [mode = title]{Proactive Recommendation in Social Networks: Steering User Interest with Causal Inference}  



%
\author[inst1]{Hang Pan}[orcid=0000-0001-8020-203X]
\ead{hungpaan@mail.ustc.edu.cn}
\author[inst1]{Shuxian Bi}
\ead{shuxianbi@mail.ustc.edu.cn}
\author[inst1]{Wenjie Wang}
\ead{wenjiewang96@gmail.com}
\cormark[1]
\author[inst3]{Haoxuan Li}
\ead{hxli@stu.pku.edu.cn}
\author[inst4]{Peng Wu}
\ead{pengwu@btbu.edu.cn}
\author[inst1]{Fuli Feng}
\ead{fulifeng93@gmail.com}
\cormark[1]

\affiliation[inst1]{organization={University of Science and Technology of China}}


\affiliation[inst3]{organization={Peking University}}

\affiliation[inst4]{organization={Beijing Technology and Business University}}













\cortext[1]{Corresponding author}



\begin{abstract}
Recommending items that solely cater to users' historical interests narrows users' horizons. 
Recent works have considered steering target users beyond their historical interests by directly adjusting items exposed to them.
However, the recommended items for direct steering might not align perfectly with the evolution of users' interests, detrimentally affecting the target users' experience.

To avoid this issue, we propose a new task named \textit{Proactive Recommendation in Social Networks} (\textbf{PRSN}) that indirectly steers users' interest by utilizing the influence of social neighbors, \ie indirect steering by adjusting the exposure of a target item to target users' neighbors.
The key to PRSN lies in answering an interventional question: \textit{what would a target user's feedback be on a target item if the item is exposed to the user's different neighbors?} 
To answer this question, we resort to causal inference and formalize PRSN as: (1) estimating the potential feedback of a user on an item,
under the network interference by the item's exposure to the user's neighbors; 
and (2) adjusting the exposure of a target item to target users' neighbors to trade off steering performance and the damage to the neighbors' experience.
To this end, we propose a \textbf{N}eighbor \textbf{I}nterference \textbf{Rec}ommendation (\textbf{NIRec}) framework with two modules: (1) an interference representation-based estimation module for modeling potential feedback; (2) a post-learning-based optimization module for adjusting a target item's exposure to trade off steering performance and the neighbors' experience through greedy search. We conduct extensive semi-simulation experiments on real-world datasets, validating the steering effectiveness of NIRec. The code
is available at \url{https://github.com/HungPaan/NIRec}.
\end{abstract}





\maketitle
\section{INTRODUCTION}\label{sec:1}
\begin{figure}[!t]
\centering
\includegraphics[scale=0.5]{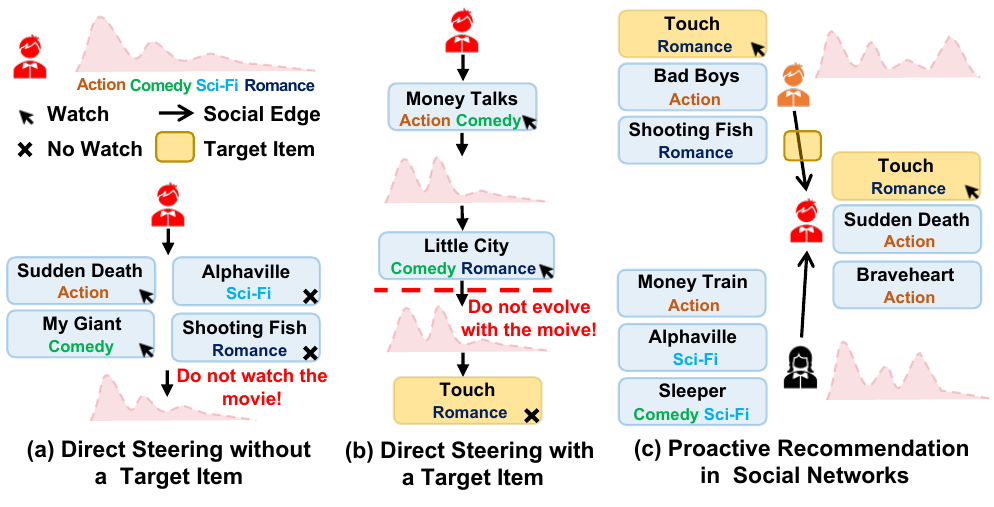}
\caption{Steering Recommendation. (a) The target user does not watch \textit{Shooting Fish} exposed to him/her directly. (b) The target user watches \textit{Little City} but the interests do not evolve with the item. (c) The target item is exposed to the target user's neighbors with the corresponding interests, leading to a successful steer.}
\label{fig:illustration}
\end{figure}

Recommender systems are widely used in various fields such as social media~\citep{zhang2023social}, content sharing~\citep{gao2023rlrec}, and e-commerce~\citep{wei2023longtail}.
A \textit{de facto} paradigm of recommender models is to learn users' interests from their historical feedback for personalized recommendation~\citep{He2020lightgcn}. 
However, reinforcing historical interests leads to filter bubbles, restricting users' exposure to diverse viewpoints and exacerbating issues like opinion polarization~\citep{gao2024filterbubble}. 
Therefore, it is crucial to build proactive recommendation that steers users beyond their historical interests. 

Existing studies on proactive recommendation directly adjust the exposed items to target users to steer user interests, roughly falling into two research lines: steering without a target item~\citep{car1998mrr,chen2018dpp, yang2023diversity} and steering with a target item~\citep{zhu2023influ,bi2024proactive}.
The former line directly exposes diverse items to target users to broaden users' horizons beyond historical interests~\citep{car1998mrr,chen2018dpp, yang2023diversity}. 
However, they overlook users’ willingness to interact with those items, potentially leading to failures in interest steering (see Figure \ref{fig:illustration} (a)). 
The latter line proactively steers a target user to interact with a target item by progressively exposing a sequence of carefully chosen items to the target user, gradually bridging the target user's current interests with the target item~\citep{zhu2023influ,bi2024proactive, lian2025itmprec, wang2024incorporate}. 
However, crafting such a sequence is significantly challenging, since it might not align perfectly with the user's interest evolution (see Figure \ref{fig:illustration} (b)). 
Misjudging the trajectory of the user's interest evolution could lead to inappropriate recommendations and detrimentally affect the user's experience.

\begin{figure}[!t]
\centering
  \includegraphics[scale=0.5]{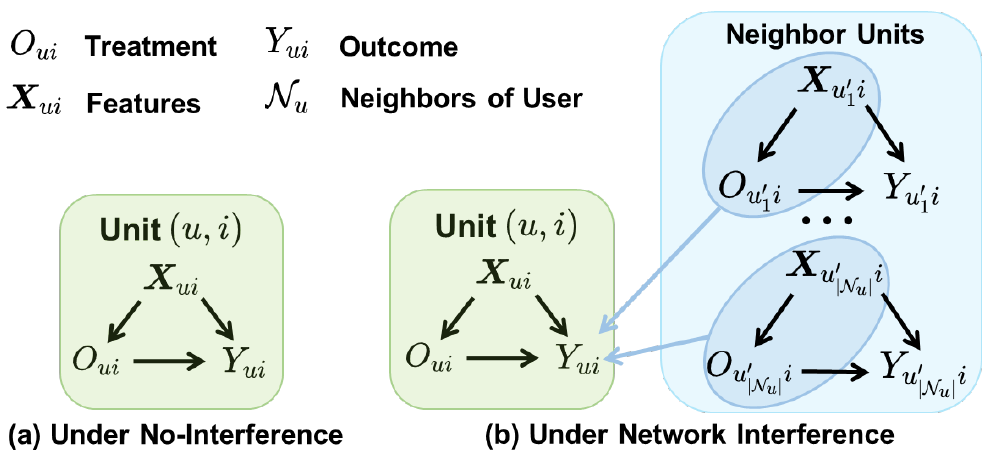}
\caption{Causal Graphs. (a) The outcome of the unit is only influenced by its features and treatment. (b) The unit's outcome is influenced by the neighbors' features and treatments.}
\label{fig:causal_graph}
\end{figure}

To avoid the problems of directly steering, we introduce a new task named \textit{Proactive Recommendation in Social Networks} (\textbf{PRSN}) --- adjusting a target item's exposure to target users' neighbors to steer their interests indirectly, given users' feedback on a target item is influenced by their neighbors on a social network~\citep{Aris2008social,lewis2012social}.
An illustration is shown in Figure \ref{fig:illustration} (c), where a target user without interest in romance movies is nudged to watch \textit{Touch} by the user's neighbors who enjoy this romance movie.
PRSN adjusts a target item's exposure to the neighbors with two objectives: 
(1) improving interaction probability between target users and the target item, and (2) limiting the damage to the neighbors' experience.
The main consideration for the second objective is the potential compromise of neighbors' experience when increasing the target item's exposure. 
For instance, to pursue steering, the target item might displace another item that better aligns with the neighbors' interests.

To achieve the objectives, the key lies in answering an interventional question: \textit{what would a target user’s feedback be on a target item if the item is exposed to the user's different neighbors?}
To answer this question, we need to resort to causal inference to estimate the potential outcomes under network interference~\citep{Ogburn2014caudiag, ma2021interference,ma2022hypergraph}.
However, traditional causal recommendation frameworks~\citep{wang2019doub, ha2024finegrained} shown in Figure \ref{fig:causal_graph}(a) typically assume no interference between units (\ie user-item pairs). Specifically, outcome $Y_{ui}$, representing whether user $u$ interacts with item $i$, is influenced only by the treatment of item exposure to $u$ and the unit's features, which can not answer the interventional question. 
To address this, we relax the no-interference assumption in Figure \ref{fig:causal_graph}(b), where new causal paths from neighbor units' treatments and features to $Y_{ui}$ indicate that the effects of neighbor units spill over, interfering with the target unit's outcome.
This interference-aware potential framework shows that the essence of achieving PRSN lies in: (1) estimating interference-aware potential outcomes for each unit; and (2) adjusting treatment values to neighbor units to trade off the probability of a positive potential outcome for the target unit and the damage to neighbors' experience, as guided by the estimated potential outcomes. 

To this end, we propose a \textbf{Ne}ighbor \textbf{I}nterference \textbf{Rec}ommend-ation  (\textbf{NIRec}) framework, which consists of two key modules: (1) an interference representation-based estimation module; and (2) a post-learning-based optimization module. 
The interference representation-based estimation module integrates a unit's individual representation with its neighbor units' interference representations to obtain the interference-aware potential outcome.
The unit's individual representation captures the user's individual interests regarding the item.
Meanwhile, the interference representations measure the interference effects of neighbor units on the target unit. 
NIRec obtains interference representations by aggregating the neighbor units' treatments and individual representations by masked graph convolution with attention. 
Given the well-trained estimation module and a pre-designed user experience cost function, the post-learning-based module greedily searches the values of neighbors' treatments to jointly maximize the positive potential outcomes probabilities of target units and minimize the decrease of the expected ratings of the neighbors. 
We conduct extensive semi-simulation experiments on real-world datasets with in-depth analysis under various settings (\eg various target users and target items), validating the steering effectiveness of NIRec.

The main contributions of this paper are as follows:
\begin{itemize}[leftmargin=*]
    \item We propose a novel \textit{Proactive Recommendation in Social Networks} (\textbf{PRSN}) task --- adjusting a target item's exposure to target users' neighbors to indirectly steer target users' interests. This task is orthogonal to existing research on direct steering, offering a new direction for steering user interest while avoiding the issues of direct steering.
    \item We design a \textbf{N}eighbor \textbf{I}nterference \textbf{Rec}ommendation (\textbf{NIRec}) framework to achieve PRSN, which leverages an interference representation-based estimation module to estimate interference-aware potential outcomes and a post-learning-based optimization module to adjust treatment values to trade off steering performance and the experience of target users' neighbors. 
    \item We conduct semi-simulation experiments on real-world datasets with in-depth analysis under various settings, verifying the effectiveness of NIRec. 
\end{itemize}

\section{RELATED WORK}
In this section, we introduce steering recommendation, causal recommendation within the potential outcome framework, and causal inference under interference, which are most relevant to our work.

\subsection{Steering Recommendation}
The first line steers users without a target item, which directly improves recommendation diversity~\citep{car1998mrr,chen2018dpp, zheng2021diversity, yang2023diversity,zhang2023diversity}.
For example, ~\citep{car1998mrr} reduces the redundancy of the recommendation list using greedy search-based re-ranking.
~\citep{zhang2023diversity} recommends multiple categories of items to users through disentangled representation learning.
While these methods broaden users' exposure to items beyond their historical interests, they do not account for users’ willingness to engage with those items, which limits their ability to effectively steer user interests.
The other line steers with a target item~\citep{zhu2023influ, bi2024proactive}.
They sequentially introduce intermediate items to users, bridging their existing interests with target items to steer their interest~\citep{zhu2023influ, bi2024proactive, wang2024incorporate, lian2025itmprec}. IRS~\citep{zhu2023influ} uses transformer-based models to create influence paths that lead users to predefined target items.  IPG~\citep{bi2024proactive} enhances adaptability by dynamically adjusting strategies based on real-time user feedback. LLM-IPP~\citep{wang2024incorporate} explores integrating large language models' reasoning and planning into proactive recommendation.
Additionally, ITMPRec~\citep{lian2025itmprec} employs a multi-round strategy to proactively steer user interests toward target items.
NIRec adjusts a target item’s exposure to a user’s neighbors to indirectly steer their interests, which is orthogonal to direct proactive recommendation, avoiding their issues while providing a new direction for steering user interests.

\subsection{Causal Recommendation}
One widely discussed task is to use a potential outcome framework to address the problem of missing not at random in recommendation~\citep{wang2019doub, guo2021mrdr, yang2023balance,zhou2023generalized, ha2024finegrained}.
These methods rely on the no-interference assumption and focus on the unbiased learning of potential outcomes for all user-item pairs with exposure treatment values of 1.
~\citep{schnabel2016recommendations,imbens2015causal,zhou2023generalized} use various strategies to estimate inverse propensity scores and reweight the biased data to align the distribution of unbiased data. 
~\citep{jiang2016doubly,wang2019doub,chen2021autodebias,guo2021mrdr,zhou2023generalized, ha2024finegrained}  incorporate loss imputation with inverse propensity scores to achieve double robustness for unbiased recommendation.
Causal recommendation also involves other problems with no-interference assumption~\citep{wu2022caurec}.
For example, ~\citep{wu2022caurec} formulates the effect of position on user feedback within a potential outcome framework by considering the item’s position in the recommendation list as a treatment.
~\citep{sato2019uplift,sato2020uplift,xie2021causalcf} consider the scenario where a user can interact with an item without recommendation and estimate the uplift of the user’s feedback from recommendation by the potential outcome framework. 
Unlike them, our work relaxes the no-interference assumption and focuses on modeling interference-aware potential outcomes.

\subsection{Causal Inference under Network Interference} 
Network interference receives widespread attention in causal inference for estimating individual treatment effects~\citep{Ogburn2014caudiag, ma2021interference,forastiere2021identification,ma2022hypergraph,jiang2022estimating,zhao2022interference}.
For example, ~\citep{Ogburn2014caudiag} distinguishes between different causal mechanisms that give rise to interference using causal graphs. 
~\citep{ma2022hypergraph} learns interference representation by hypergraph convolution with attention.
Considering interference, ~\citep{ma2021interference} maximizes the utility on the entire network by optimizing all units' treatments in the population on the network.
Our work differs from these studies in two key ways: (1) constructing target user-item pairs and adjusting the neighbors' treatment values to achieve our task for those target pairs; and (2) considering user experience under treatment adjustment.


\section{TASK FORMALIZATION}
In this section, we present traditional recommendation from the view of causal inference without interference. 
Then we introduce PRSN and formalize PRSN's objectives from the perspective of causal inference under network interference. 

\subsection{A Causal View of Traditional Recommendation}
We have a classic causal graph as shown in Figure \ref{fig:causal_graph}(a), which depicts causal relations among different variables in recommendation.
$u$ and $i$ are a user and an item in user set $\mathcal{U}$ and item set $\mathcal{I}$, respectively. 
User-item pair $(u,i)$ is regarded as a unit. 
The exposure treatment of unit $(u,i)$ indicates whether item $i$ is exposed to user $u$, which is a binary random variable denoted as $O_{ui}$. 
In particular, if item $i$ is exposed to user $u$, $O_{ui}=1$; otherwise $O_{ui}=0$. 
The feature vector of unit $(u,i)$ contains features of user $u$ and item $i$, such as user ID and item ID, which is denoted as $\bm{X}_{ui}$. 
For unit $(u,i)$, binary potential outcome $Y_{ui}(\bm{O}=\bm{o})$ refers to user $u$'s potential feedback (\eg click) on item $i$ given treatment values of $\bm{o}$, where $\bm{O}$ is a treatment vector containing treatments of all units.
Under the no-interference assumption that potential outcomes of a unit are not influenced by treatments of other units, we obtain the widely studied potential outcome $Y_{ui}(O_{ui}=1)$ for unit $(u,i)$ in recommendation.

In the factual world, we can only observe one potential outcome of unit $(u,i)$, which is called the outcome of unit $(u,i)$ and denoted as $Y_{ui}$. 
For example, if unit $(u,i)$ was assigned with a treatment value of $O_{ui}=0$, potential outcome $Y_{ui}(O_{ui}=1)$ cannot be observed. 
Correspondingly, the observed feedback data in the factual world is denoted as $\mathcal{D}=\{\bm{x}_{ui},o_{ui}, y_{ui}|o_{ui}=1, u\in \mathcal{U},i\in \mathcal{I}\}$. 
The objective of traditional recommendation is to: (1) estimate the probability of positive potential outcome $P(Y_{ui}(O_{ui}=1)=1|\bm{X}_{ui}=\bm{x}_{ui})$ for each unit by the observed feedback data; and (2) recommend items with the highest probabilities to cater to users.

\subsection{Formalization of PRSN with Causal Inference \\ Under Network Interference} \label{sec:forma}
Considering that users' feedback on a target item is influenced by their neighbors in social networks~\citep{Aris2008social,lewis2012social}, we believe that the interests of a user's neighbors in the exposed items can be propagated through social networks.
Therefore, we propose PRSN,  which indirectly nudges target users to interact with a target item by adjusting this item's exposure to the social neighbors of target users. 
Given that increasing the target item's exposure may damage the neighbors' experience, PRSN has two objectives: (1) improving the interaction probability between the target users and the target item; and (2) limiting the damage to the neighbors' experience. 
The key to PRSN is to answer an interventional question: \textit{what would a target user’s feedback on a target item be if the item is exposed to the user's different neighbors?} 
To answer this question, we resort to causal inference under network interference~\citep{Ogburn2014caudiag,ma2021interference,ma2022hypergraph}.

Let $\mathcal{S}=\{(u,u^\prime)\mid u\in \mathcal{U}, u^\prime \in \mathcal{U}, s_{uu^\prime}=1\}$ denote a directed social network, 
where $s_{uu^\prime}=1$ indicates that user $u$ trusts user $u^\prime$.
We define user $u$'s neighbors as the users connected with user $u$ in social network $\mathcal{S}$, denoted as $\mathcal{N}_u=\{u^\prime|s_{uu^\prime}=1, u^\prime \in \mathcal{U}\}$.
Similarly, we define the neighbor units of unit $(u,i)$ as the user-item pairs composed of the neighbors of user $u$ and item $i$, which is denoted as $(\mathcal{N}_u,i) = \{(u^\prime,i)\mid u^\prime \in \mathcal{N}_u\}$.
Treatments of unit $(u,i)$'s neighbor units are denoted as $\bm{O}_{\mathcal{N}_{u}i}=[O_{u^\prime_1i}, O_{u^\prime_2i}, \cdots, O_{u^\prime_{|\mathcal{N}_{u}|}i}]^\intercal$.
Furthermore, $\mathcal{U}^\ast \subset  \mathcal{U}$ and $i^\ast \in \mathcal{I}$ are target users and a target item, respectively. $\mathcal{U}^\ast$'s neighbors are users connected to them in the social network while excluding those who belong to $\mathcal{U}^\ast$, which is denoted as $\mathcal{N}_{\mathcal{U}^\ast}=\{u^\prime|s_{uu^\prime}=1, u\in \mathcal{U}^\ast, u^\prime \notin \mathcal{U}^\ast\}$.
With the presence of neighbor influence in social networks, it is necessary to be aware of two types of new causal paths related to network interference (see Figure \ref{fig:causal_graph}(b)).
In particular, $\bm{X}_{u^\prime i},O_{u^\prime i} \rightarrow Y_{ui}$ depicts that the effect of neighbor $(u^\prime, i)$'s treatment $O_{u^\prime i}$ and features $\bm{X}_{u^\prime i}$ spills over to unit $(u,i)$, interfering with unit $(u,i)$'s outcome $Y_{ui}$.
It relaxes the no-interference assumption of traditional causal recommendation.

Considering interference, we have $Y_{ui}(\bm{O})=Y_{ui}(O_{ui},\bm{O}_{\mathcal{N}_ui})$. Correspondingly, the potential outcome of unit $(u,i)$ under a treatment value of $O_{ui}=1$ can be converted as $Y_{ui}(O_{ui}=1, \bm{O}_{\mathcal{N}_{u}i})$. Under such a formulation, we formalize PRSN as the following two points: 
\begin{itemize}[leftmargin=*]
    \item Estimating the probability of positive potential outcomes under varying interference (\ie $\bm{o}_{\mathcal{N}_ui}$) for each unit:
        \begin{equation}\label{eq:po}
            P\left(Y_{ui}\left(O_{ui}=1, \bm{O}_{\mathcal{N}_{u}i}=\bm{o}_{\mathcal{N}_{u}i}\right)=1\mid \bm{X}_{ui}\right),
        \end{equation}
    which answers the interventional question.
    \item Balancing the sum of probabilities of positive potential outcomes of target units $\sum_{u^{\ast}\in \mathcal{U^\ast} }P(Y_{u^{\ast}i^\ast}(\bm{O}_{\mathcal{U}^\ast i^\ast} =\bm{1}, \bm{O}_{\mathcal{N}_{\mathcal{U}^\ast }i^\ast}=\bm{o}_{\mathcal{N}_{\mathcal{U}^\ast}i^\ast})=1\mid \bm{X}_{u^{\ast}i^{\ast}}=\bm{x}_{u^{\ast}i^{\ast}})$
    and the damage to the neighbors $\sum_{u^\prime \in \mathcal{N}_{\mathcal{U}^\ast}}c(o_{u^\prime i^\ast})$ by adjusting $\bm{o}_{\mathcal{N}_{\mathcal{U}^\ast}i^{\ast}}$.
    Here $c(o_{u^\prime i^\ast})$ is a cost function measuring the damage of $o_{u^\prime i^\ast}$ to neighbor $u^\prime$ through the decrease in $u^{\prime}$'s expected rating, with further details provided in Section~\ref{sec:pos}.

\end{itemize}

\section{METHOD}\label{sec:method}
\begin{figure}[!t]
\centering
  \includegraphics[scale=0.55]{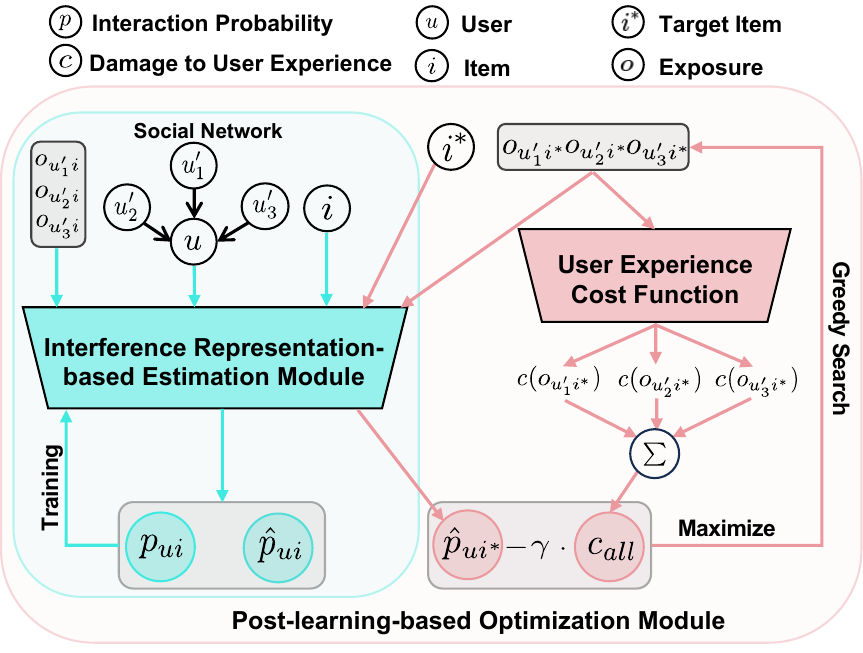}
\caption{NIRec. Initially, train an interference representation-based estimation module for potential outcome estimation. Then, optimize the exposure of the target item to the user's neighbors, given the well-trained estimation module and a predesigned cost function.}
\label{fig:framework}
\end{figure}

In this section, we introduce NIRec as shown in Figure \ref{fig:framework}, which consists of an interference representation-based potential outcome estimation module and a post-learning-based treatment value optimization module to achieve PRSN.

\subsection{Identification}
We identify interference-aware potential outcomes with standard assumptions in the causal inference literature~\citep{Ogburn2014caudiag, ma2021interference,ma2022hypergraph}.

\newtheorem{assumption}{Assumption}
\begin{assumption}\label{ass:inter}
There exists an interference representation function $\phi(\cdot)$ sufficiently capturing information of $(u,i)$'s neighbors' treatment values $\bm{o}_{\mathcal{N}_ui}$ and features $\bm{\mathrm{X}}_{\mathcal{N}_{u}i}$.
If $\phi(\cdot)$ is determined, $(u,i)$'s potential outcome is determined. 
Formally, for any $(u,i)$, $\bm{o}^1_{\mathcal{N}_{u}i}$, $\bm{o}^2_{\mathcal{N}_{u}i}$, $\bm{\mathrm{X}}^1_{\mathcal{N}_{u}i}$, and $\bm{\mathrm{X}}^2_{\mathcal{N}_{u}i}$, if $\phi(\bm{o}^1_{\mathcal{N}_{u}i}, \bm{\mathrm{X}}^1_{\mathcal{N}_{u}i})=\phi(\bm{o}^2_{\mathcal{N}_{u}i}, \bm{\mathrm{X}}^2_{\mathcal{N}_{u}i})$, we have:
\begin{equation}
    Y_{ui}\left(O_{ui}=1, \bm{O}_{\mathcal{N}_{u}i}=\bm{o}^1_{\mathcal{N}_{u}i}\right)=Y_{ui}\left(O_{ui}=1, \bm{O}_{\mathcal{N}_{u}i}=\bm{o}^2_{\mathcal{N}_{u}i}\right).
\end{equation}
In other words, given $\bm{g}_{ui}=\phi(\bm{o}_{\mathcal{N}_{u}i}, \bm{\mathrm{X}}_{\mathcal{N}_{u}i})$, we have:
\begin{equation}
Y_{ui}\left(O_{ui}=1, \bm{O}_{\mathcal{N}_{u}i}=\bm{o}_{\mathcal{N}_{u}i}\right)=Y_{ui}\left(O_{ui}=1, \bm{G}_{ui}=\bm{g}_{ui}\right).
\end{equation}
\end{assumption}

\begin{assumption}\label{ass:indepen}
Given the unit's features, the potential outcomes are independent of its own treatment and the interference representation.
Formally,  we have:
\begin{equation}
Y_{ui}\left(O_{ui}=1,\bm{G}_{ui}=\bm{g}_{ui}\right)\perp \left(O_{ui}, \bm{G}_{ui}\right) \mid \bm{X}_{ui}.
\end{equation}

\end{assumption}

\begin{assumption}\label{ass:consis}
The potential outcome of a unit under a specific treatment value and a specific interference representation is the same as the observed outcome when that unit receives that treatment value and interference representation. Formally, if $O_{ui}=1$ and $\bm{G}_{ui}=\bm{g}_{ui}$, we have:
\begin{equation}
Y_{ui}=Y_{ui}\left(O_{ui}=1, \bm{G}_{ui}=\bm{g}_{ui}\right).
\end{equation}
\end{assumption}

According to the assumptions, the identification of  $P(Y_{ui}(O_{ui}=1, \bm{O}_{\mathcal{N}_{u}i}=\bm{o}_{\mathcal{N}_{u}i})=1\mid  \bm{X}_{ui}=\bm{x}_{ui})$ can be proved as follows:
\begin{equation}
    \begin{split}
&P\left(Y_{ui}\left(O_{ui}=1, \bm{O}_{\mathcal{N}_{u}i}=\bm{o}_{\mathcal{N}_{u}i}\right)=1\mid  \bm{X}_{ui}=\bm{x}_{ui}\right)\\
\overset{a}{=}&P\left(Y_{ui}\left(O_{ui}=1, \bm{G}_{ui}=\bm{g}_{ui}\right)=1\mid  \bm{X}_{ui}=\bm{x}_{ui}\right)\\
\overset{b}{=}&P\left(Y_{ui}\left(O_{ui}=1, \bm{G}_{ui}=\bm{g}_{ui}\right)=1 \mid  \bm{X}_{ui}=\bm{x}_{ui}, {O}_{ui}=1, \bm{G}_{ui}=\bm{g}_{ui}\right)\\
\overset{c}{=}&P\left(Y_{ui}=1 \mid  \bm{X}_{ui}=\bm{x}_{ui}, {O}_{ui}=1, \bm{G}_{ui}=\bm{g}_{ui}\right).
    \end{split}
    \end{equation}
Equations (a), (b), and (c) are based on Assumptions \ref{ass:inter}, \ref{ass:indepen}, and \ref{ass:consis}, respectively. 
It is worth noting that the three standard assumptions commonly employed in causal inference are equally applicable in recommendation. Assumption \ref{ass:inter} benefits significantly from the strong representational capabilities of graph convolution in recommendation~\citep{He2020lightgcn}, which can be adapted to capture information regarding the neighbors' features and the exposure of the item to the neighbors. Assumptions \ref{ass:indepen} and  \ref{ass:consis} are adaptations of conventional assumptions in the traditional causal recommendation~\citep{wu2022caurec}, tailored to account for the presence of interference.

\subsection{Interference Representation-based Estimation \\ Module}
We propose an interference representation-based estimation module consisting of a masked graph convolution with attention. 
The interference representation-based estimation module first models unit $(u,i)$'s individual representation to capture user $u$'s individual interests regarding item $i$ as follows:
\begin{equation}
    \bm{e}_{ui} = \bm{e}_u\odot \bm{e}_i,
\end{equation}
where  $\bm{e}_u$ and $\bm{e}_i$ are learnable embeddings for user $u$ and item $i$, respectively,
which can be modeled by previous recommender models such as Matrix Factorization (MF)~\citep{koren2009mf} and LightGCN~\citep{He2020lightgcn}.

In social networks, when $u$'s neighbor $u^\prime$ is exposed to item $i$, $u^\prime$ may be inclined to share its interests $\bm{e}_{u^\prime i}$ with  $u$. 
Moreover, the social closeness between different neighbors and a user is not uniform, resulting in differing degrees of influence on the user's interests in the item.
Therefore, the interference representation-based estimation module aggregates unit $(u,i)$'s neighbors' treatment values $\bm{o}_{\mathcal{N}_{u}i}$ and user $u$'s neighbors' individual interests in item $i$ as follows:
\begin{equation}\label{eq:aggregate}
    \bm{g}_{ui}=\bm{\mathrm{W}}\sum_{u^\prime \in \mathcal{N}_u}o_{u^\prime i}a_{uu^\prime}\bm{e}_{u^\prime i},
\end{equation}
where $\mathbf{W}$ is a learnable matrix and $a_{uu^\prime}$ is the attention weight between user $u$ and $u$'s neighbor $u^\prime$.
The attention captures the importance of the social edge between user $u$ and $u$'s neighbor $u^\prime$ \wrt user $u$’s feedback:
\begin{equation}\label{eq:attention}a_{uu^\prime}=\frac{\exp(\bm{w}^\intercal [\bm{e}_{u}\parallel \bm{e}_{u^\prime}])}{\sum_{u^{\prime \prime}\in \mathcal{N}_u} \exp(\bm{w}^\intercal[\bm{e}_{u}\parallel \bm{e}_{u^{\prime \prime}}])},\end{equation}
where $\bm{w}$ is a learnable weight vector and $[\cdot||\cdot]$ denotes the concatenation of two vectors.

Concatenating unit $(u,i)$'s individual representation and interference representation as $[\bm{e}_{ui}\parallel \bm{g}_{ui}]$, we can capture user $u$'s final inclination towards interacting with the item $i$.
Any classifier can be used to model the interference potential outcome of unit $(u,i)$ (\eg a linear classifier). 
In this work, we follow MF and model $P(Y_{ui}(O_{ui}=1, \bm{O}_{\mathcal{N}_{u}i}=\bm{o}_{\mathcal{N}_{u}i})=1\mid \bm{X}_{ui}=\bm{x}_{ui})$ as:
\begin{equation}
\sigma\left(\bm{1}^\intercal[\bm{e}_{ui}\parallel \bm{g}_{ui}]\right),
\end{equation}
where $\bm{1}$ is a vector where each element's value is 1 and $\sigma(\cdot)$ is the sigmoid function. 
In estimating interference-aware potential outcome, we can specify item $i$'s exposure $\bm{o}_{\mathcal{N}_{u}i}$ to user $u$'s neighbors by observed data $\mathcal{D}$.
The modeled interference-aware potential outcomes are denoted as $\hat{p}_{ui}$.
We optimize the above learnable parameters by minimizing the following loss:
\begin{equation}
\frac{1}{|\mathcal{D}|}\sum_{(\bm{x}_{ui}, o_{ui}, y_{ui})\in \mathcal{D}}l(\hat{p}_{ui}, y_{ui}),
\end{equation}
where $l(\cdot, \cdot)$ is the cross-entropy loss.

\subsection{Post-Learning-based Optimization Module}\label{sec:pos}
We propose a post-learning-based optimization module to achieve the second point in Section \ref{sec:forma}, which consists of: (1) estimating the sum of probabilities of positive potential outcomes of target units under interference; (2) designing a cost function to measure the damage to the neighbors resulting from the target item's exposure; and (3) adjusting the treatment values of the neighbors to trade off steering performance and the damage to the neighbors.

\subsubsection{Estimation for target units}
Given the well-trained interference representation-based estimation module, we can estimate the sum of probabilities of positive potential outcomes of target units under interference $\sum_{u^{\ast}\in \mathcal{U^\ast} }P(Y_{u^{\ast}i^\ast}(\bm{O}_{\mathcal{U}^\ast i^\ast} =\bm{1}, \bm{O}_{\mathcal{N}_{\mathcal{U}^\ast }i^\ast}=\bm{o}_{\mathcal{N}_{\mathcal{U}^\ast}i^\ast})=1\mid \bm{X}_{u^{\ast}i^{\ast}}=\bm{x}_{u^{\ast}i^{\ast}})$. Although $Y_{u^\ast i^{\ast}}( \bm{O}_{\mathcal{U}^\ast i^\ast} =\bm{1}, \bm{O}_{\mathcal{N}_{\mathcal{U}^\ast }i^\ast}=\bm{o}_{\mathcal{N}_{\mathcal{U}^\ast}i^\ast})$ contains treatment values of units other than unit $(u^\ast,i^\ast)$ and its neighbor units, we can also use the estimation module, because we have $Y_{u^\ast i^\ast}(\bm{O})=Y_{u^\ast i^\ast}(O_{u^\ast i^\ast},\bm{O}_{\mathcal{N}_{u^\ast}{i^\ast}})$.
\subsubsection{Design of cost function}\label{sec:cost}
We utilize an existing recommender model as the user experience model\footnote{In practice, we train a LightGCN~\citep{He2020lightgcn}, a representative recommender model, as a user experience model, since we currently lack access to the recommendation platforms' proprietary recommender models designed for user experience.} and based on this, we design a cost function. 
The cost function measures the damage to a neighbor by measuring the decrease in the sum of ratings for items in the adjusted recommendation list from NIRec compared to the sum of ratings in the original recommendation list from the user experience model, both ratings as assigned by the user experience model.
In particular,  given the user experience model, we obtain: (1) an item list with the top k highest ratings to neighbor $u^\prime$; and (2) an exposure treatment value $o^{ex}_{u^\prime i^\ast}$ indicating whether $i^\ast$ is in the top k list ($o^{ex}_{u^\prime i^\ast}=1$) or not ($o^{ex}_{u^\prime i^\ast}=0$).
From the user experience model's perspective, if we replace an item in the top k item list, it will damage $u^\prime$'s experience. Therefore, we design cost function $c(\cdot)$ as follows:
\begin{itemize}[leftmargin=*]
    \item If NIRec does not expose item $i^\ast$ to neighbor $u^\prime$, but the user experience model does, we need to choose another item to replace target item $i^\ast$ in the top $k$ items list.
    The resulting decrease of the rating indicates the damage to $u^\prime$'s experience.
    Formally, if $o_{u^\prime i^\ast}=0$ and $o^{ex}_{u^\prime i^\ast}=1$, we have $c(o_{u^\prime i^\ast})=\min_{i} r^{ex}_{u^\prime i^\ast}-r^{ex}_{u^\prime i}$.
    The replacement is the $k+1$-th highest-rated item of the user experience model for $u^\prime$.
    \item If NIRec exposes item $i^\ast$ to neighbor $u^\prime$, but the user experience model does not, we need to replace an item in the top $k$ item list with $i^\ast$.
    The rating decrease also indicates the damage to $u^\prime$'s experience.
    Formally, if $o_{u^\prime i^\ast}=1$ and $o^{ex}_{u^\prime i^\ast}=0$, we have $c(o_{u^\prime i^\ast})=\min_{i}r^{ex}_{u^\prime i}-r^{ex}_{u^\prime i^\ast} $. 
     The replacement is the $k$-th item in the top $k$ list.
     \item If both NIRec and the user experience model expose target item $i^\ast$ to neighbor $u^\prime$, or if neither exposes it, there is no damage to neighbor $u^\prime$'s experience.
     Formally, if $o_{u^\prime i^\ast}=o^{ex}_{u^\prime i^\ast}$, we have $c(o_{u^\prime i^\ast})=0$.
\end{itemize}

\subsubsection{Treatment values optimization}
We trade off the improvement of steering performance and the damage to their neighbors. This is achieved by maximizing a bi-objective function over $\bm{o}_{\mathcal{N}_{\mathcal{U}^\ast}i^\ast}$. The two-objective function is defined as:
\begin{equation}\label{eq:trade}
\begin{split}
        &\sum_{u^{\ast}\in \mathcal{U^\ast} }P(Y_{u^{\ast}i^\ast}(\bm{O}_{\mathcal{U}^\ast i^\ast} =\bm{1}, \bm{O}_{\mathcal{N}_{\mathcal{U}^\ast }i^\ast}=\bm{o}_{\mathcal{N}_{\mathcal{U}^\ast}i^\ast})=1\mid \bm{X}_{u^{\ast}i^{\ast}}=\bm{x}_{u^{\ast}i^{\ast}})\\
        &- \gamma \sum_{u^\prime \in \mathcal{N}_{\mathcal{U}^\ast}}c\left(o_{u^\prime i^\ast}\right).
\end{split}
\end{equation}
To efficiently solve this NP-hard maximization problem, we use a greedy search algorithm. Initially, we set $\bm{O}_{\mathcal{N}_{\mathcal{U}^\ast}i^\ast}$ to $\bm{o}^{ex}_{\mathcal{N}_{\mathcal{U}^\ast}i^\ast}$. Then, we iteratively identify and reassign the element within the vector $\bm{o}_{\mathcal{N}_{\mathcal{U}^\ast}i^\ast}$ that yields the maximal increase in the value of the objective function (\ref{eq:trade}). This process is repeated until the objective function's value plateaus, indicating that no further gains can be achieved. Once an element in the $\bm{O}_{\mathcal{N}_{\mathcal{U}^\ast}i^\ast}$ is reassigned, it is excluded from further consideration in subsequent iterations.

By incrementally adjusting the trade-off coefficient $\gamma$ within a predefined range using a set step size, we can obtain solutions reflecting different trade-offs.
The final solution can be selected to align with the specific requirements.
For example, under a given constraint on the damage to the neighbors, we can choose the solution with the highest potential outcome for superior user interest steering. 

\subsubsection{Scalability}
NIRec comprises two modules: the interference representation-based estimation module and the post-learning-based optimization module. The interference representation-based estimation module has a time complexity of \( O\left( \left|\mathcal{D}\right| \left( k \, d + d^{2} \right) \right) \), where \( \left|\mathcal{D}\right| \) is the size of the training data, \( k \) is the average number of social neighbors per user, and \( d \) is the embedding dimension. Its computational cost scales linearly with both the dataset size and the density of the social network. Given the precomputed interaction probability of the target users for the target item under the initial exposure, the post-learning-based optimization module has a time complexity of \( O\left( k \left( Ft - \frac{t(t-1)}{2} \right) \right) \), where \( F \) is the number of the neighbors of the target group, \( k \) is the number of target users influenced by each neighbor, and \( t \) is the number of iterations required for the optimization objective to converge (worst-case \( t = F \), but in practice \( t < F \)). Consequently, this module's worst-case complexity is \( O\left( F^{2} k \right) \) and its best-case complexity is \( O\left( Fk \right) \). If, in reality, the growth rate of social network density is lower than that of the network's scale, NIRec demonstrates good scalability. In the worst-case scenario, however, as the network scale increases, \( \left|\mathcal{D}\right| \), \( k \), and \( F \) all increase linearly, limiting NIRec's scalability. Future work will focus on further optimizing NIRec's scalability.

\subsubsection{Discussion}
Our framework can be extended to other types of networks where indirect influence might also play a role. The key lies in constructing causal graphs corresponding to the networks that relax interference assumptions. For example, in an item-item similarity network, exposing items related to the target item to the target user will influence the user's feedback on the target item. The corresponding causal graph will include new causal paths: $\bm{X}_{u i^\prime},O_{u i^\prime} \rightarrow Y_{ui}$, where $i^\prime$ is the neighbor item of item $i$ in the item-item similarity network. 
We can employ heterogeneous graph neural networks to simultaneously model the interference brought by different networks. We will explore this further in future work.
\section{EXPERIMENT SETUP}
\begin{table}[!t]
    \centering
    \caption{The statistics of Ciao, Epinions, Filmtrust.}
\begin{tabular}{c|ccc}
    \hline
    Dataset & Ciao & Epinions & Filmtrust\\
    \hline
    \#Users & 2,541 & 12,168 & 1,208 \\
    \#Items & 3,019 & 11,283 & 406 \\
    \#Interactions & 71,670 & 352,598 & 31,668 \\
    \#Social Edges & 48,282 & 123,187 & 1,270 \\
    Edge Density&0.748\%&0.083\%&0.087\%\\ 
    \#Interfered User-Item Pairs & 179,727 & 55,029 & 13,992 \\
    \hline
\end{tabular}
    \label{tab:statistics}
\end{table}
Since no offline datasets provide users' feedback under varying neighbor interference, we propose a framework for user feedback simulation. 
It is noteworthy that semi-simulation is a viable and commonly practiced experimental method in direct steering~\citep{zhu2023influ, bi2024proactive, wang2024incorporate, lian2025itmprec} and for recommendation tasks in interactive evaluation settings~\citep{sato2019uplift, gao2024filterbubble,sato2020uplift,gao2023rlrec}.
Following the user feedback simulation, we present how to select target users and items for PRSN. 
Finally, we design evaluation metrics and baselines.
\subsection{Real-World Datasets}\label{app:data}
We use two product review datasets, Ciao and Epinions~\citep{tang2012ciao1, tang2012ciao2}, as well as a movie review dataset Filmtrust~\citep{guo2013film}, for simulating users' feedback under varying neighbor interference, which results in three user feedback simulations.
These datasets contain user IDs, item IDs, ratings (from 1 to 5), and a directed social network with trust relations. We use the 10-core setting for Ciao and Epinions and the 5-core setting\footnote{The number of items in Filmtrust is on the order of 100, which is relatively small.} for Filmtrust. The statistics of the datasets are shown in Table \ref{tab:statistics}.

\subsection{User Feedback Simulation}
We design three user feedback simulations based on the above three real-world datasets and name them as Sim Ciao, Sim Epinions, and Sim Filmtrust, respectively.

\subsubsection{Representation Simulation}
We simulate representations for both users and items to generate user feedback under neighbor interference. 
To this end, we train an MF model using the observed ratings to predict and complete missing ratings in the user-item matrix.
However, the model always falls into overestimation on missing ratings~\citep{guo2021mrdr}.
To this end, we follow~\citep{guo2021mrdr} to align the distribution of the ratings in the completed matrix with that of ratings in the random exposure dataset (\eg Yahoo!R3~\citep{marlin2012yahoo}), thereby enhancing the reliability of our simulations.
Thereafter, we use all ratings to train a new MF model to obtain our simulated user representation $\bm{\mathrm{E}}_\mathcal{U}^\ast$ and item representation $\bm{\mathrm{E}}_\mathcal{I}^\ast$.

\subsubsection{User Feedback Generation Function}\label{sec:ufgf}
According to standard potential outcome simulation in causal inference~\citep{ma2021interference,ma2022hypergraph}, we design a user feedback generation function considering neighbor interference. 
Given $\bm{\mathrm{E}}_\mathcal{U}^\ast$  and $\bm{\mathrm{E}}_\mathcal{I}^\ast$, we set the positive potential outcome probability of unit $(u,i)$ as:
\begin{equation}\label{groundtruth}
\begin{split}
\sigma\left(\eta\left(\bm{e}^\ast_u, \bm{e}^\ast_i\right)+\beta \Delta\left(\bm{\mathrm{E}}^\ast_{\mathcal{N}_u}, \bm{e}^\ast_u,\bm{e}^\ast_i, \bm{o}_{\mathcal{N}_{u}i}\right)\right),
\end{split}
\end{equation}
where $\eta(\bm{e}^\ast_u, \bm{e}^\ast_i)$ is user $u$'s individual interests in interacting with item $i$, which can be shortened to $\eta_{ui}$ for brevity. 
$\Delta(\bm{\mathrm{E}}^\ast_{\mathcal{N}_u}, \bm{e}^\ast_u,\bm{e}^\ast_i, \bm{o}_{\mathcal{N}_{u}i})$ is the neighbors' interference effect, which is shortened to $\Delta_{ui}$. 
$\beta$ is the coefficient that controls the strength of the interference. 
In particular, we design $\eta_{ui}$  and $\Delta_{ui}$ under a linear setting:
\begin{equation}
    \begin{split}
    \eta_{ui}&=\bm{e}^\ast_u \cdot \bm{e}^\ast_i - 3,\\   
    \Delta_{ui}&=\frac{1}{\sqrt{\left|\mathcal{N}_u\right|} }\sum_{u^\prime\in\mathcal{N}_u}{o_{u^\prime i}\sigma\left(\bm{e}_u^\ast\cdot \bm{e}_{u^\prime}^\ast\right)\eta_{u^\prime i}}.
    \end{split}
\end{equation}
Here, we consider a user to like an item if the corresponding rating is above 3; otherwise, the user is considered to dislike the item, given that the range of ratings fitted by MF for simulation is from 1 to 5. If the user's neighbor's rating of the item is higher than 3, the influence on the user is positive; otherwise, it is negative.

\subsubsection{Semi-Synthetic Datasets}\label{app:semi-data}
We generate synthetic datasets with user feedback to train our interference representation-based estimation module. 
We adjust the propensity for observing the feedback to ensure that the generated data aligns as closely as possible with the characteristics of real-world data.

The propensity for observing the feedback in the factual world has three characteristics: (1) the propensity is highly related to the user's individual interests; (2) the propensity has randomness; (3) the propensities \wrt two user-item pairs connected on a social network are related. 
Therefore, we set the propensity for observing the feedback as $\alpha_1 p_1 + \alpha_2 p_2 + \alpha_3 p_3$, where $p_1$, $p_2$, and $p_3$ are the propensity \wrt the user's individual interests, randomness, and the number of \textit{the interfered user-item pairs} (\ie connected user-item pairs that both feedback is observed $\sum_{(u,i)\in\mathcal{D}}|\{(u^\prime, i)\mid u^\prime \in \mathcal{N}_{u}, (u^\prime, i)\in\mathcal{D}\}|$), respectively.
$\alpha_1$, $\alpha_2$, $\alpha_3$, $p_1$, $p_2$, and $p_3$ are constrained to $0 \leq \alpha_1, \alpha_2, \alpha_3, p_1, p_2, p_3 \leq 1$ and $\alpha_1+\alpha_2+\alpha_3=1$.
We adjust $|\mathcal{D}|$ and $\alpha_3$ to ensure the semi-synthetic dataset matches the real-world dataset in terms of the number of observed user-item pairs with feedback and the number of the interfered user-item pairs. 
\subsubsection{Discussion for Semi-Simulation}
Our semi-simulation aims to address the absence of user feedback under varying neighbor interference in offline datasets. 
It consists of two components: the mechanism of neighbor influence on user feedback in social networks and the user feedback dataset used for training.
For the influence mechanism of neighbors on user feedback, we rely on a well-known and widely accepted assumption: users' feedback is influenced by their neighbors' interests and the strength of social relationships~\citep{lewis2012social}.  
Specifically, we assume that recommending the target item to the neighbors of the target user can influence the target user's feedback on that item, and this influence can be either positive or negative.
We implement this mechanism based on real-world offline datasets, where the user/item representation simulation follows the design in~\citep{guo2021mrdr}, and the user feedback generation function follows~\citep{ma2021interference,ma2022hypergraph}. For the user feedback set used for training, we align the semi-synthetic data with real-world offline datasets as much as possible, including the propensity of observing feedback, randomness, data sparsity, and the number of connected users exposed to the same item. These efforts ensure that our semi-simulation reflects real-world scenarios as closely as possible.
Nonetheless, our current semi-simulation lacks a mechanism to model how influence propagates through multi-hop neighbors. We plan to address these issues in future work.

\subsection{Target Users and Target Items Selection}
We determine the target users and target items for the PRSN task based on the following rules: 

\textbf{The selection criteria for target user groups} are as follows:(1) each user in the target group must have at least one neighbor who is interested in the target item, providing an opportunity for the target user to be steered by their neighbors\footnote{PRSN assumes that a user cannot be steered to interact with a target item if none of this user's neighbors like it.}; and (2) individual interest regarding the target item of the user in the group is below a threshold $t\le1$, \ie $\eta_{u^\ast i^\ast}\le t$, indicating the user is not very interested in the target item and can be steered.
\textbf{The selection criterion for a target item} is straightforward: once a target item is chosen, the number of qualified target users must exceed a predetermined size. For example, suppose the predetermined size of a target user group is 50, there must be a minimum of 50 users in the entire user set who meet the selection criteria of target users for the target item. 

In practice, we select $1,000$ target items and then select a predetermined size of target user group for each target item\footnote{For Filmtrust with items fewer than $1,000$, we choose all possible target items.}. 
We conduct experiments on the aforementioned target user groups and items in turn and average the experiment results to ensure the reliability and effectiveness of our conclusions.
The above selection process is implemented by rejection sampling without replacement.
Taking the selection of a target item as an example, we randomly sample an item in the entire item set and determine whether to reject the item according to the criterion for selecting a target item.

\subsection{Evaluation Metrics}
We design two metrics to evaluate model performance for PRSN.
\begin{itemize}[leftmargin=*]
\item \textbf{Improvement of Interaction Probability (IoIP)} measures the improvement in interaction probability following steering, compared to the interaction probability \wrt the user's individual interests.
Formally, we have:
\begin{equation}
   \text{IoIP} \triangleq \frac{1}{|\mathcal{U}^\ast|} \sum_{u^\ast\in \mathcal{U}^\ast}\frac{\sigma(\eta_{u^\ast i^\ast}+\beta \Delta_{u^\ast i^\ast})-\sigma(\eta_{u^\ast i^\ast})}{\sigma(\eta_{u^\ast i^\ast})}.
\end{equation}
A higher IoIP means better steering performance.
\item \textbf{Damage to Neighbors' Experience (DtNE)} measures the damage to the neighbors' experience resulting from adjusting the target item's exposure to them.
Formally, we have:
\begin{equation}
\text{DtNE} \triangleq \sum_{u^\prime \in \mathcal{N}_{\mathcal{U}^\ast}}c(o_{u^\prime i^\ast}).
\end{equation}
A smaller DtNE means a better user experience from the perspective of the selected user experience model.
\end{itemize}

IoIP directly corresponds to our first goal of PRSN — whether steering raises a target user’s probability of interacting with the target item. Formally, it computes the average relative uplift in interaction probability, normalized by each user’s initial interaction probability without steering. IoIP enjoys the following properties: (1) relativity: by dividing by the initial probability, it fairly compares gains across users with different initial interaction probabilities; (2) boundedness: it has a lower bound (-1) and no upper bound—this means it can strongly reward increasing interaction probabilities for items that initially have low probabilities, precisely matching the proactive recommendation objective of steering users beyond their historical interests; (3) monotonicity: any positive steering adjustment yields a non‑negative increase.
DtNE directly corresponds to our second goal—how much steering damages the experience of the neighbors of the target users.  It calculates the total damage across all neighbors. DtNE enjoys the following properties: (1) cumulative measurement: it sums the total damage to all neighbors’ experience, enabling holistic auditing; (2) non‑negativity: it is always greater than or equal to 0, so zero steering yields zero damage; (3) flexibility: by selecting different cost functions, one can measure the damage to the neighbors' experience from different perspectives. The cost function used in this study is defined in Section \ref{sec:cost}.

\subsection{Baselines}
Existing recommender models are not suitable for PRSN as they do not model the effect of the target item's exposure to the target user's neighbors on the target user's feedback.
We are the first to do this task.
Therefore, we propose two heuristic strategies to adapt existing recommender models for PRSN as follows:
\begin{itemize}[leftmargin=*]
    \item \textbf{Direct. }Take the same direction of neighbor interference as the user feedback generation function \ref{groundtruth}. For items that are liked, neighbors will have a positive effect on the target user, and vice versa. Formally, the strategy is maximizing the following objective over $\bm{o}_{\mathcal{N}_u{i^\ast}}$:
\begin{equation}\label{eq:baseline1}
\begin{split}
     \sum_{u^\prime\in \mathcal{N}_{\mathcal{U}^\ast}} o_{u^\prime i^\ast}\left(\sigma\left(\bm{e}^b_{u^\prime}\cdot \bm{e}^b_{i^\ast}\right)-0.5\right) - \gamma \sum_{u^\prime \in \mathcal{N}_{\mathcal{U}^\ast}}c(o_{u^\prime i^\ast}),
\end{split}
\end{equation}
where $\bm{e}^b_u$ and $\bm{e}^b_i$ are the selected recommender model's user representation and item representation, respectively.
\item \textbf{Sim. }Take a consistent assumption of neighbor interference with user feedback generation function \ref{groundtruth}. Formally, the strategy is maximizing the following objective over $\bm{o}_{\mathcal{N}_u{i^\ast}}$:
\begin{equation}\label{eq:baseline2}
\begin{split}
&\sum_{u^\ast\in \mathcal{U^\ast} }\sigma\left(\eta\left(\bm{e}^b_{u^{\ast}}, \bm{e}^b_{i^\ast}\right)+\beta \Delta\left(\bm{\mathrm{E}}^b_{\mathcal{N}_{u^{\ast}}}, \bm{e}^b_{u^\ast},\bm{e}^b_{i^\ast}, \bm{o}_{\mathcal{N}_{u^{\ast}}{i^\ast}}\right)\right) \\ &- \gamma \sum_{u^\prime \in \mathcal{N}_{\mathcal{U}^\ast}}c\left(o_{u^\prime i^\ast}\right),
\end{split}
\end{equation}
where $\bm{\mathrm{E}}^b_{\mathcal{N}_u}$ is user representation matrix for $\mathcal{N}_u$.
\end{itemize}

Accordingly, we design 6 baselines based on MF, LightGCN~\citep{He2020lightgcn}, and DiffNet~\citep{wu2019diffnet}, which are Direct MF, Sim MF, Direct LightGCN, Sim LightGCN, Direct DiffNet, and Sim  DiffNet.
MF is the most classic and fundamental model in recommendation, which can be viewed as a variant of NIRec removing the aggregation of the item's exposure to the user's neighbors and the neighbors' features. 
LightGCN is a representative model for graph-based recommendation.
DiffNet is the representative model that aggregates the features of the user's neighbors, which can be viewed as a variant of NIRec removing the aggregation of the item's exposure to the user's neighbors.
\textbf{Note that social recommendation methods like DiffNet leverage information from social networks to better align recommendations with users' interests. However, they fail to steer target users beyond their historical interest because they neglect the effect of exposing an item to neighbors on a user (rely on no-interference assumption).
}

\section{EXPERIMENTS}

\begin{table*}[ht]
\centering
\caption{Overall performance without damage constraints across different target user group sizes given $\beta=10$ and $t=1$. The bold and underlined fonts indicate the highest and second-highest IoIP other than Oracle.}
\label{tab:overall}
\resizebox{\textwidth}{!}{ 
\begin{tabular}{cc|cccc|cccc|cccc}
    \hline
     & Simulation & \multicolumn{4}{c|}{Sim Ciao} & \multicolumn{4}{c|}{Sim Epinions} & \multicolumn{4}{c}{Sim Filmtrust}  \\
    \hline
     & Group Size   & 1  & 50  & 75  & 100  & 1  & 50  & 75  & 100 & 1  & 25  & 50  & 75       \\
    \hline
\multirow{6}{*}{IoIP} 
& Direct MF   & -0.047  & -0.176  & -0.198  & -0.217  & 0.338  & 0.195  & 0.157  & 0.156  & 0.347  & 0.190  & 0.187  & 0.204     \\
& Sim MF   & -0.039  & -0.175  & -0.198  & -0.218  & 0.344  & 0.197  & 0.159  & 0.157  & 0.346  & 0.190  & 0.187  & 0.204     \\
& Direct DiffNet   & 0.085  & -0.056  & \underline{-0.085}  & \underline{-0.115}  & 0.183  & 0.093  & 0.065  & 0.049  & 0.305  & 0.155  & 0.152  & 0.170     \\
& Sim DiffNet   & \underline{0.106}  & \underline{-0.056}  & -0.090  & -0.120  & 0.212  & 0.103  & 0.072  & 0.055  & 0.304  & 0.155  & 0.151  & 0.169     \\
& Direct LightGCN   & -0.001  & -0.127  & -0.148  & -0.170  & 0.362  & 0.208  & 0.174  & 0.172  & \underline{0.348}  & \underline{0.198}  & \underline{0.193}  & \underline{0.204}     \\
& Sim LightGCN   & 0.018  & -0.126  & -0.151  & -0.173  & \underline{0.376}  & \underline{0.213}  & \underline{0.178}  & \underline{0.175}  & 0.344  & 0.196  & 0.191  & 0.202     \\
& NIRec   & \textbf{0.226}  & \textbf{0.059}  & \textbf{0.044}  & \textbf{0.016}  & \textbf{0.477}  & \textbf{0.297}  & \textbf{0.267}  & \textbf{0.267}  & \textbf{0.468}  & \textbf{0.283}  & \textbf{0.279}  & \textbf{ 0.292}     \\
& Oracle   & 0.655  & 0.442  & 0.405  & 0.352  & 0.669  & 0.524  & 0.492  & 0.477  & 0.548  & 0.355  & 0.339  & 0.341     \\
\hline \hline
\multirow{6}{*}{DtNE} 
& Direct MF   & 2.7  & 86.7  & 109.0  & 123.4  & 1.8  & 56.9  & 76.5  & 94.7  & 0.6  & 8.9  & 15.3  & 20.6     \\
& Sim MF   & 2.2  & 73.9  & 94.7  & 107.5  & 1.6  & 50.6  & 68.4  & 85.6  & 0.6  & 8.7  & 15.0  & 20.2     \\
& Direct DiffNet   & 2.3  & 69.9  & 86.6  & 97.6  & 2.2  & 66.9  & 87.7  & 107.0  & 0.6  & 9.5  & 16.4  & 22.0     \\
& Sim DiffNet   & 1.4  & 45.7  & 58.1  & 65.1  & 1.6  & 48.5  & 64.0  & 78.9  & 0.6  & 8.8  & 15.3  & 20.7     \\
& Direct LightGCN   & 2.3  & 72.3  & 87.7  & 99.7  & 1.8  & 56.4  & 74.5  & 94.1  & 0.5  & 8.4  & 14.1  & 18.9     \\
& Sim LightGCN   & 1.6  & 54.9  & 68.1  & 78.2  & 1.4  & 44.5  & 60.4  & 77.3  & 0.5  & 7.7  & 13.0  & 17.5     \\
& NIRec   & 2.5  & 78.7  & 100.4  & 112.1  & 1.2  & 43.7  & 58.0  & 73.2  & 0.8  & 12.7  & 20.5  & 25.8     \\
& Oracle   & 3.1  & 78.8  & 99.0  & 110.6  & 1.7  & 51.8  & 66.1  & 81.5  & 0.8  & 13.1  & 21.1  & 27.0     \\
    \hline
\end{tabular}
} 

\end{table*}

In this section, we conduct experiments to answer the following research questions:
\begin{itemize}[leftmargin=*]
\item \textbf{RQ1}: Does NIRec effectively improve the interaction probability between the target user group and the target item?
\item \textbf{RQ2}: How does neighbor interference strength coefficient $\beta$ influence steering performance?
\item \textbf{RQ3}: How does individual interest threshold $t$ influence steering performance?
\item \textbf{RQ4}: Can the effectiveness of steering user interests be further improved by adjusting the target user group?
\end{itemize}
To reveal the ideal results,  we view the user feedback generation function with the simulated representation as a method for PRSN, named Oracle. 
For brevity, certain experiments' results about Epinions and Filmtrust are shown in \ref{app:epinions}.

\subsection{Overall Performance (RQ1)}
\subsubsection{Steering without Damage Constraints}Table \ref{tab:overall} shows the results of steering target user groups of different sizes to interact with the target item without damage constraints ($\gamma=0$) given $t=1$ and $\beta=10$.
From the table, we have the following observations:
\begin{itemize}[leftmargin=*]
\item NIRec outperforms all baselines across all cases. This superiority underscores the effectiveness of NIRec and highlights the significance of estimating interference-aware potential outcomes—a factor overlooked by all baselines.
\item Compared to DiffNet-based baselines, MF-based and LightGCN-based baselines achieve higher IoIP in Epinions and Filmtrust, which have lower social edge density (0.083\% and 0.087\%). DiffNet-based baselines attain higher IoIP in Ciao, featuring higher social edge density (0.748\%).
This finding suggests that these baselines lack consistent performance across diverse datasets, whereas NIRec demonstrates robust, generalized steering capabilities.
\item With comparable IoIP, Direct-based baselines have higher DtNE than Sim-based baselines. This is because Direct-based baselines indiscriminately select all neighbors who like the target item, disregarding the fact that optimal steering requires only a subset of these neighbors. This inefficiency indicates that existing recommender models fall short of the ideal recommendation for PRSN.
\end{itemize}

\begin{figure*}[!t]
\centering
\includegraphics[scale=0.135]{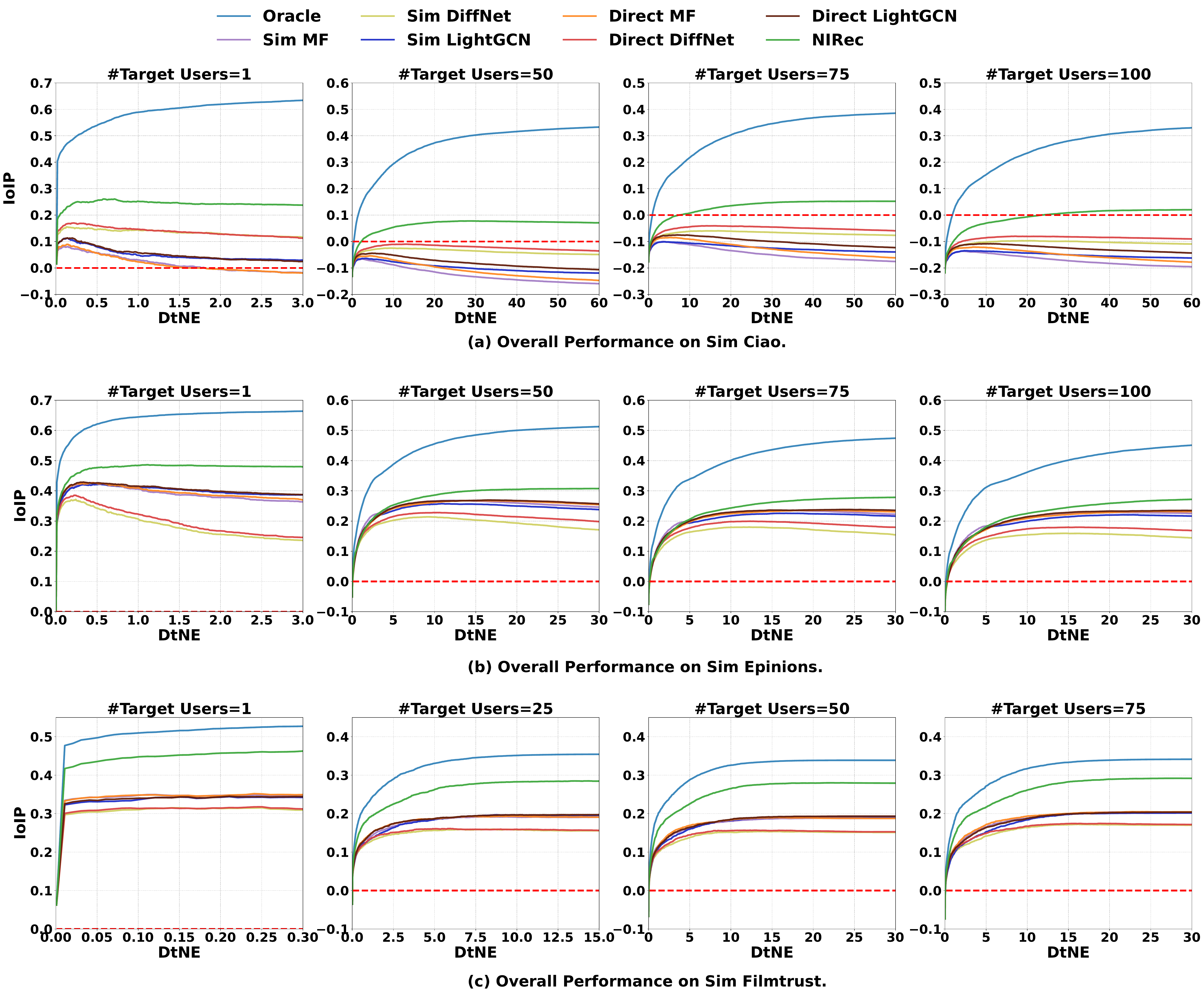}
\caption{DtNE-IoIP curves across different target user group sizes given $\beta=10$ and $t=1$.}
\label{fig:steering_overall}
\end{figure*}
\subsubsection{Steering with Damage Constraints}
Figure \ref{fig:steering_overall} shows DtNE-IoIP curves across different target user group sizes given $\beta=10$ and $t=1$. The group sizes are 25, 50, and 75 on Sim Filmtrust because it has fewer users.
From the figure, we observe the following:
\begin{itemize}[leftmargin=*]
\item Without proactive steering (DtNE = 0), IoIP of target user groups with sizes of 50, 75, and 100 (25, 50, and 75 on Sim Filmtrust) are negative. 
This occurs because the individual interests of the selected target users in the target item are below the threshold ($t \le 1$), leading to scenarios where these users negatively influence each other’s interest in the target item through the social network.
These results reveal that merely exposing the target item to the target user groups not only fails to improve the interaction probability but may also have a detrimental effect. Therefore, proactive steering is necessary, given the existence of neighbor interference.
\item NIRec outperforms all baselines across all damage constraints and target user group sizes. Furthermore, its IoIP is consistently higher than IoIP resulting from only directly exposing the target item to the target user group. These findings confirm the efficacy of NIRec.
\end{itemize}

\begin{figure*}[!t]
\centering
\includegraphics[scale=0.135]{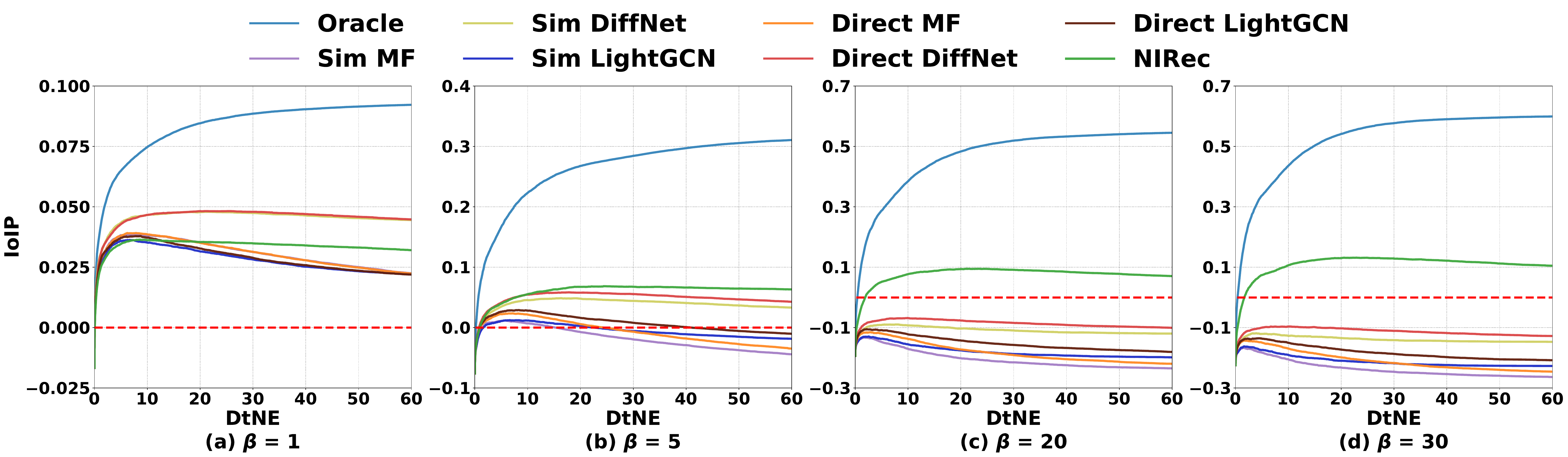}
\caption{DtNE-IoIP curves across different interference strength given a group size of 50 and $t=1$ in Sim Ciao.}
\label{fig:beta_ana}
\end{figure*}

\begin{figure}[!t]
\centering
\includegraphics[scale=0.130]{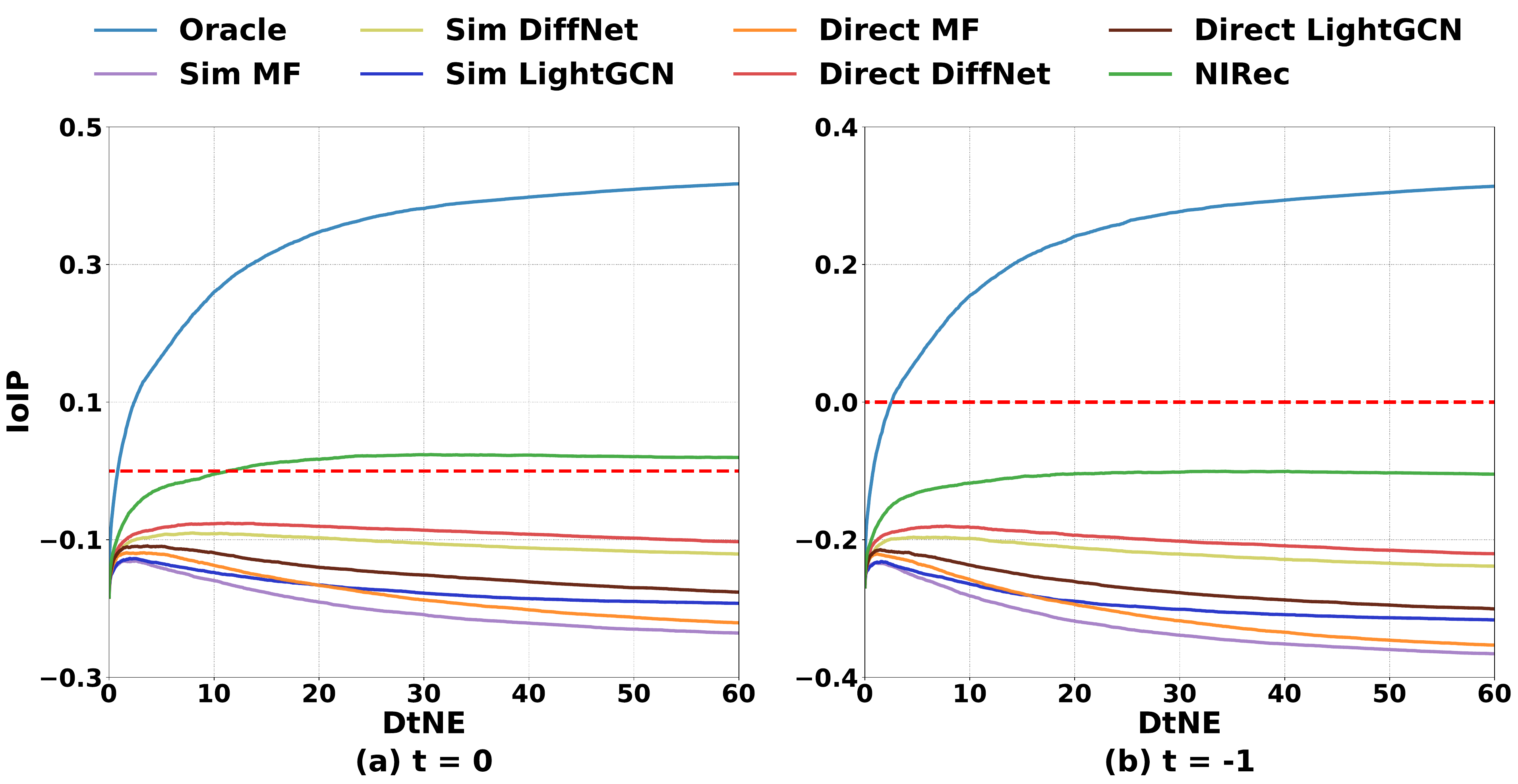}
\caption{DtNE-IoIP curves across individual interests threshold given a group size of 50 and $\beta=10$ in Sim Ciao.}
\label{fig:threshold_ana}
\end{figure}

\begin{figure}[!t]
\centering
\includegraphics[scale=0.120]{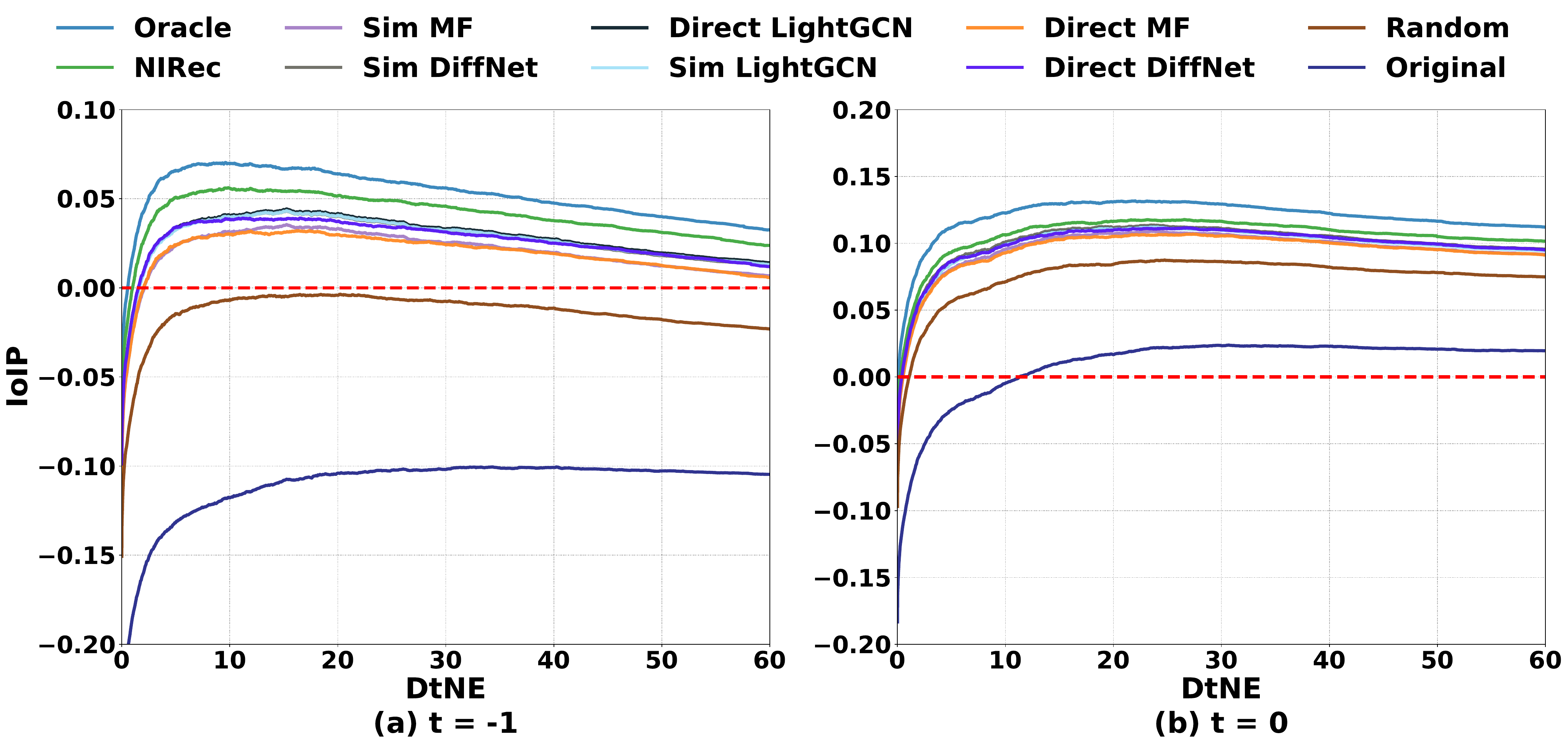}
\caption{NIRec's DtNE-IoIP curves \wrt the original target user group and the adjusted target user groups by different methods across different thresholds given a group size of 50 and $\beta=10$ in Sim Ciao.}
\label{fig:population_ana}
\end{figure}

\subsection{In-Depth Analysis (RQ2, RQ3, RQ4)}
\subsubsection{Influence of Interference Strength Coefficient (RQ2)}
Figure \ref{fig:beta_ana} presents DtNE-IoIP curves across different interference strength coefficients $\beta$, given a group size of 50 and $t=1$ on Sim Ciao.
First, we observe: (1) given a suitable damage constraint, NIRec's IoIP remains positive across all $\beta$ values; (2) all baselines have negative IoIP at $\beta=10$ (Figure \ref{fig:steering_overall}), $\beta=20$, and $\beta=30$; and (3) NIRec outperforms all baselines when $\beta \neq 1$. These results validate the necessity of estimating interference-aware potential outcomes, especially in cases of high neighbor interference strength.
Second, we observe that NIRec is inferior to Direct DiffNet but still has positive IoIP at $\beta=1$. The reason is that the interference strength is too weak to enable good estimations of the potential outcomes.

\subsubsection{Influence of Individual Interests Threshold (RQ3)}
Figure \ref{fig:threshold_ana} presents DtNE-IoIP curves across different individual interest thresholds given a group size of 50 and $\beta=10$ on Sim Ciao.
In Figure \ref{fig:threshold_ana}, we make three key observations: 
\begin{itemize}[leftmargin=*]
\item The IoIP of all methods decreases as $t$ becomes smaller, indicating that the lower the target group's interest in the target item, the harder it is to steer users within the target group.
\item All methods, other than Oracle, exhibit very low IoIP, especially at $t=-1$, where the IoIP is negative across all damage constraints. This indicates that when the interference strength estimation is insufficiently accurate, we cannot simultaneously steer all users within the target user group who have extremely low interest in the target item. 
The reason is that when target users have very low interest in the target item, the negative effects associated with the target item, propagated through the social network among users within the group, may outweigh the positive effects of steering efforts.

\item NIRec outperforms all baselines at $t=0$ and $t=-1$, demonstrating its ability to better mitigate the negative impact on the target group.
\end{itemize}

\subsubsection{Target User Group Adjustment Study (RQ4)}
Figure \ref{fig:population_ana} shows NIRec's DtNE-IoIP curves \wrt the original and adjusted target groups using different methods across different thresholds given a group size of 50 and $\beta=10$ in Sim Ciao.
In particular, we have two strategies to select users in the target user group: (1) use different methods to respectively select users in the target user group that produce the highest negative impact on interaction probability; and (2) randomly select users in the target user group.
The adjusting process is: (1) selecting users in the target user group based on the above strategies; (2) cutting the social edges from the selected users to other users in the group; and (3) comparing IoIP of NIRec on different adjusted target user groups to IoIP of NIRec on the original group.
We observe: (1) NIRec's IoIP significantly improves after adjusting the target user group using any strategy and any method, indicating that the adjustment of cutting edges in the target user group is important for improving steering performance; (2) adjustments by strategy 1 are better than strategy 2, showing all methods can effectively identify target users' negative impact in the target group rather than only decreasing the number of social edges; and (3) adjustment by NIRec works best, validating its effectiveness.
In real-world applications, cutting social edges is challenging. 
Nonetheless, these results provide two valuable insights for better steering performance:  (1) construct a group where there are fewer social edges between users within the target user group, which leads to a lower negative impact produced in the group; and (2) first remove a select number of users producing the highest negative impact, then focus on steering the remaining users in the target user group, which results in improved performance.

\section{CONCLUSION}

In this paper, we explored the capability of recommendation to steer users beyond their historical interests. 
We proposed a new task PRSN, which adjusts the exposure of a target item to a target user's neighbors to nudge the target user toward interacting with that item. 
To implement PRSN, we proposed a framework named NIRec, which (1) introduces an interference representation-based estimation module to estimate the potential user feedback on a target item interfered with by the user’s neighbors, and (2) utilizes a post-learning-based optimization module to adjust the neighbors' exposure to trade off steering performance and the damage to the neighbors’ experience. 
Extensive experiments on real-world datasets validate the steering effectiveness of NIRec.

This work presents a novel neighbor steering recommendation task, opening up many new research directions. 
One direction for future work is to model the interference from multi-hop neighbors on target users' feedback. 
Besides, it is valuable to extend our methods to multiple rounds of recommendation to capture the neighbors' long-term influence on the target users' interests. 













\printcredits
\appendix

\section{Experiments}\label{app:epinions}
\begin{figure*}[!t]
\centering
\includegraphics[scale=0.135]{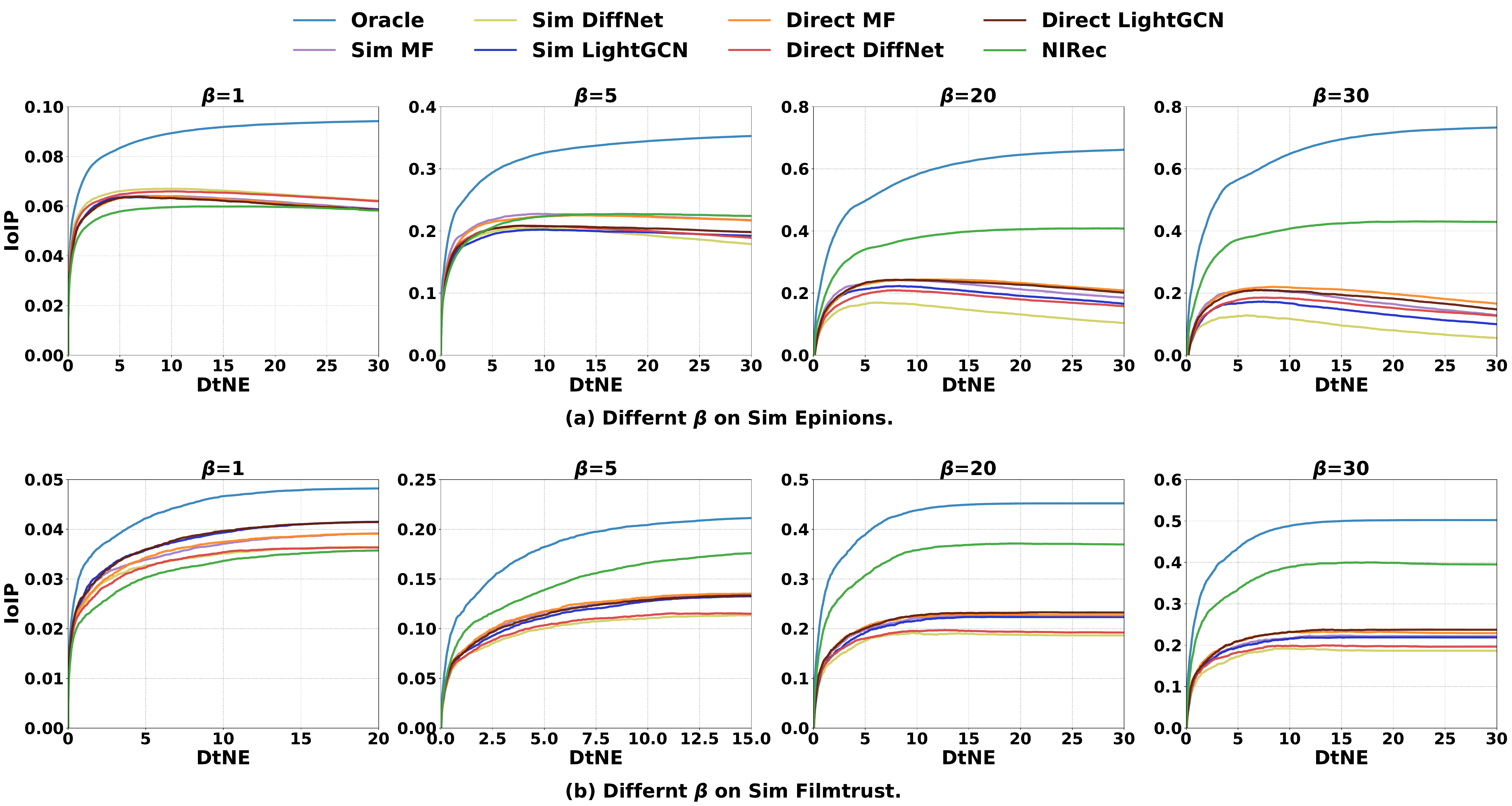}
\caption{DtNE-IoIP curves across different interference strength coefficients given a group size of 50 and $t=1$ in Sim Epinions and Sim Filmtrust.}
\label{fig:epi_beta_ana_ef}
\end{figure*}

\begin{figure*}[!t]
\centering
\includegraphics[scale=0.135]{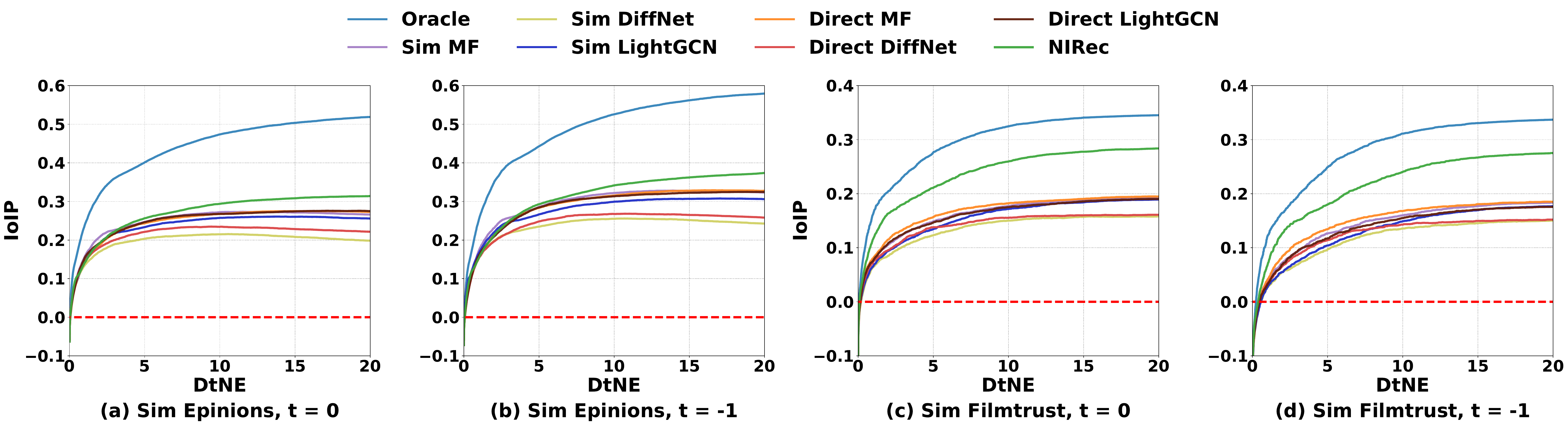}
\caption{DtNE-IoIP curves across individual interests threshold given a group size of 50 and $\beta=10$ in Sim Epinions and Sim Filmtrust.}
\label{fig:epi_threshold_ana_ef}
\end{figure*}

\begin{figure*}[!t]
\centering
\includegraphics[scale=0.135]{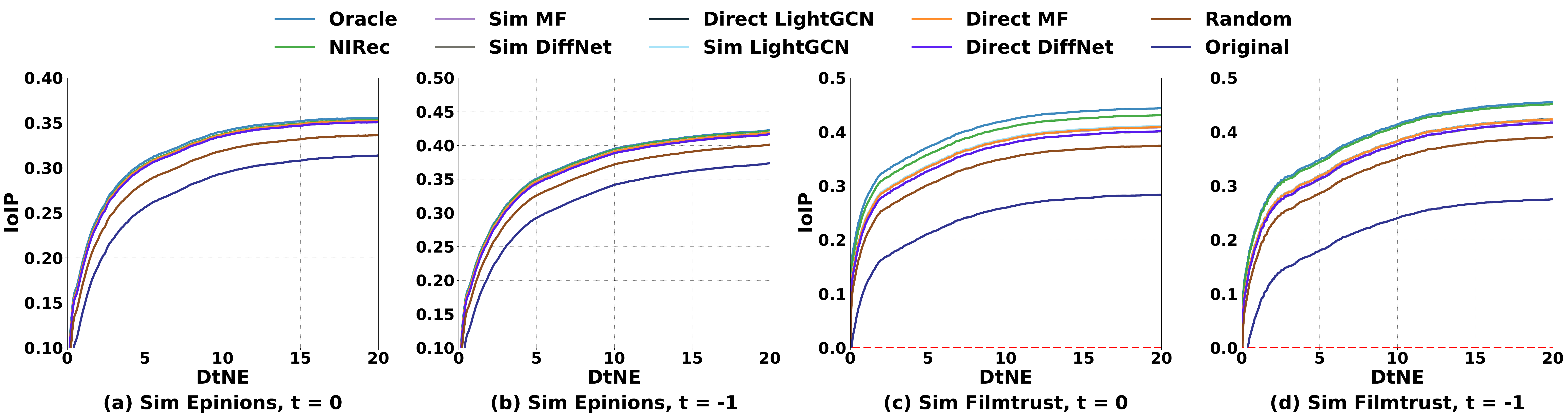}
\caption{NIRec's DtNE-IoIP curves \wrt the original target user group and the adjusted target user groups by different methods across different thresholds given a group size of 50 and $\beta=10$ in Sim Epinions and Sim Filmtrust.}
\label{fig:epi_population_ana_ef}
\end{figure*}

In contrast to the results obtained on Sim Ciao, experiments conducted in Sim Epinions and Sim Filmtrust (shown in Figure \ref{fig:epi_beta_ana_ef}, Figure \ref{fig:epi_threshold_ana_ef}, and Figure \ref{fig:epi_population_ana_ef}) often exhibit positive IoIP. 
We attribute this phenomenon to the lower social edge density within Epinions and Filmtrust, which consequently leads to lower edge density among users within the target user group and a weaker negative influence propagated among users within the target group through the social network.
This experimental observation also conveniently corroborates the insight from the target user group adjustment study --- constructing a group where there are fewer social edges between users within the group will lead to a lower negative impact produced in the group and achieve better steering performance.

\bibliographystyle{cas-model2-names-issue}

\bibliography{cas-refs}



\end{document}